\documentclass[letterpaper,twocolumn,10pt]{article}
\usepackage{graphics}
\usepackage{url}
\usepackage{verbatim}
\usepackage{myusenix}
\author{Petros Maniatis \hspace{4mm} Mary Baker\\ {\em Computer
Science Department, Stanford University}\\ {\em Stanford, CA 94305,
USA}\\
{\tt \{maniatis,mgbaker\}@cs.stanford.edu}\\
\url{http://identiscape.stanford.edu/}} % end author

\date{}
\begin{document}

\title{\bf Secure History Preservation Through\\
Timeline Entanglement}

\maketitle

%For the first page - it's separate from the rest in terms of page style.
\thispagestyle{empty}

\begin{abstract}

\small A \emph{secure timeline} is a tamper-evident historic record
of the states through which a system goes throughout its operational
history.  Secure timelines can help us reason about the temporal
ordering of system states in a provable manner.  We extend secure
timelines to encompass multiple, mutually distrustful services,
using \emph{timeline entanglement}.  Timeline entanglement
associates disparate timelines maintained at independent systems, by
linking undeniably the past of one timeline to the future of
another.  Timeline entanglement is a sound method to map a time step
in the history of one service onto the timeline of another, and
helps clients of entangled services to get persistent temporal
proofs for services rendered that survive the demise or
non-cooperation of the originating service.  In this paper we
present the design and implementation of \emph{Timeweave}, our
service development framework for timeline entanglement based on two
novel disk-based authenticated data structures. We evaluate
Timeweave's performance characteristics and show that it can be
efficiently deployed in a loosely-coupled distributed system of a
few hundred services with overhead of roughly 2-8\% of the
processing resources of a PC-grade system.

\end{abstract}

\section{\label{sec:Introduction}Introduction}

A large portion of the functionality offered by current commercial
``secure'' or ``trusted'' on-line services focuses on the here and
now: certification authorities certify that a public signature
verification key belongs to a named signer, secure file systems
vouch that the file with which they answer a lookup query is the one
originally stored, and trusted third parties guarantee that they do
whatever they are trusted to do when they do it.

The concept of \emph{history} has received considerably less
attention in systems and security research.  What did the
certification authority certify a year ago, and which file did the
secure file system return to a given query last week?

Interest in such questions is fueled by more than just curiosity.
Consider a scenario where Alice, a certified accountant, consults
confidential documents supplied by a business manager at client
company Norne, Inc.\ so as to prepare a financial report on behalf
of the company for the Securities and Exchange Commission (SEC).
If, in the future, the SEC questions Alice's integrity, accusing her
of having used old, obsolete financial information to prepare her
report, Alice might have to prove to the SEC exactly what
information she had received from Norne, Inc.\ before preparing her
report.  To do that, she would have to rely on authentic historic
data about documents and communication exchanges between herself and
Norne, on the authentic, relative and absolute timing of those
exchanges, perhaps even on the contents of the business agreement
between herself and the company at the time.  Especially if the
company maliciously chooses to tamper with or even erase its local
records to repudiate potential transgressions, Alice would be able
to redeem herself only by providing undeniable proof that at the
time in question, Norne, Inc.\ did in fact present her with the
documents it now denies.

Besides this basic problem, many other peripheral problems lurk:
what if Norne, Inc.\ no longer exists when Alice has to account for
her actions?  What if Alice and the SEC belong to different trust
domains, i.e., have different certification authorities or different
secure time stamping services?

In this work we formulate the concept of secure timelines based on
traditional time stamping~\cite{Haber1991,Benaloh1991} and
authenticated dictionaries~\cite{Buldas1998,Goodrich2001}
(Section~\ref{sec:Timelines}).  Secure timelines allow the
maintenance of a persistent, authenticated record of the sequence of
states that an accountable service takes during its lifetime.

Furthermore, we describe a technique called \emph{timeline
entanglement} for building a single, common tamper-evident history
for multiple mutually distrustful entities
(Section~\ref{sec:Entanglement}).  First, timeline entanglement
enables the temporal correlation of independent histories, thereby
yielding a single timeline that encompasses events on independent
systems.  This correlation can be verified independently in the
trust domain of each participant, albeit with some loss of temporal
resolution.  Second, it allows clients to preserve the provability
of temporal relationships among system states, even when the systems
whose states are in question no longer participate in the
collective, or are no longer in existence.

We then present \emph{Timeweave}, our prototype framework for the
development of loosely-coupled distributed systems of accountable
services that uses timeline entanglement to protect historic
integrity (Section~\ref{sec:Implementation}).  We describe novel,
scalable algorithms to maintain secure timelines for extended time
periods and for very large data collections.  Finally, we evaluate
the performance characteristics of Timeweave in
Section~\ref{sec:Evaluation} and show that it efficiently supports
large-sized groups of frequently entangled services---up to several
hundred---with maintenance overhead that does not surpass 2-8\% of
the computational resources of a PC-grade server.

\section{\label{sec:Background}Background}

In this work we draw on results from research on secure time
stamping and authenticated dictionaries.  The main inspiration
behind our approach comes from Lamport's classic logical clock
paradigm~\cite{Lamport1978}.

\subsection{\label{sec:Timestamping}Secure Time Stamping}

In secure time stamping, it is the responsibility of a centralized,
trusted third party, the \emph{Time Stamping Service} (TSS), to
maintain a temporal ordering of submission among digital documents.
As documents or document digests are submitted to it, the TSS links
them in a tamper-evident chain of authenticators, using a one-way
hash function, and distributes portions of the chain and of the
authenticators to its clients.  Given the last authenticator in the
chain it is impossible for anyone, including the TSS, to insert a
document previously unseen in the middle of the chain unobserved,
without significant collusion, and without finding a second pre-image
for the hash function used~\cite{Haber1991}.

Benaloh and de Mare~\cite{Benaloh1991} describe synchronous,
broadcast-based time stamping schemes where no central TSS is
required, and introduce the concept of a time stamping \emph{round}.
All documents time stamped during a round are organized in a data
structure, flat or hierarchical, and yield a collective digest that
can be used to represent all the documents of the entire round, in a
tamper-evident manner; given the digest, the existence of exactly
the documents inside the data structure can be proved succinctly, and
any document outside the data structure can be proved not to be
there.

Buldas et al.~\cite{Buldas1998} extend previous work by
significantly diminishing the need to trust the TSS.  They also
introduce efficient schemes for maintaining relative temporal
orderings of digital artifacts with logarithmic complexity in the
total number of artifacts.  A large, concurrent project towards the
full specification of a time stamping service is described by
Quisquater et al.~\cite{Quisquater1999}.

Ansper et al.~\cite{Ansper2001} discuss time stamping service
availability, and suggest a scheme similar to consensus in a
replicated system to allow for fault-tolerant time stamping.

\subsection{\label{sec:AuthenticatedDictionaries}Authenticated
Dictionaries}

Authenticated dictionaries are data structures that operate as
tamper-evident indices for a dynamic data set.  They help compute
and maintain a one-way digest of the data set, such that using this
digest and a succinct proof, the existence or non-existence of any
element in the set can be proved, without considering the whole set.

The first such authenticated dictionary is considered to be an
unanticipated use of Merkle's hash trees~\cite{Merkle1980}, a
digital signature scheme.  Hash trees are binary trees in whose
leaves the data set elements are placed.  Each leaf node is labeled
with the hash of the contained data element and each interior node
is labeled with a hash of the concatenated labels of its children.
The label of the root node is a tamper-evident digest for the entire
data set.  The existence proof for an element in the tree consists
of the necessary information to derive the root hash from the
element in question; specifically, the proof consists of all labels
and locations (left or right) of all siblings of nodes on the path
from the element to the tree root.

Tree-based authenticated dictionaries reminiscent of Merkle's hash
trees have been most notably used for the distribution of
certificate revocation records, first by Kocher~\cite{Kocher1998},
and then in an incrementally updatable version by Naor and
Nissim~\cite{Naor1998}.  Buldas et al.\ have obviated the need for
trusting the dictionary maintainer to keep the dictionary sorted, by
introducing the \emph{authenticated search
tree}~\cite{Buldas2000,Buldas2002}. Authenticated search trees are
like hash trees, but all nodes, leaves and internal nodes alike,
contain data set elements.  The label of the node is a hash not only
of the labels of its children, but also of the element of the node.
Existence proofs contain node elements in addition to nodes'
siblings' labels on the path from the element in question to the
root.  In this manner, an existence proof follows the same path that
the tree maintainer must take to find a sought element; as a result,
clients need not unconditionally trust that the tree maintainer
keeps the tree sorted, since given a root hash, there is a unique
descent path that follows the standard traversal of search trees
towards any single element.

Authenticated dictionaries have also been proposed based on
different data structures.  Buldas et al.~\cite{Buldas1998} describe
several tree-like ``binary linking schemes.'' Goodrich et
al.~\cite{Goodrich2001} propose an authenticated skip list that
relies on commutative hashing.

In the recent literature, the maintenance of authenticated but
persistent dynamic sets~\cite[p. 294]{CLRv2} has received some
attention.  Persistent dynamic sets allow modifications of the
elements in the set, but maintain enough information to recreate any
prior version of the set.  Anagnostopoulos et
al.~\cite{Anagnostopoulos2001} propose and implement persistent
authenticated skip lists, where not only older versions of the skip
list are available, but they are each, by themselves, an
authenticated dictionary.  In the same work, and also in work by
Maniatis and Baker~\cite{Maniatis2002}, persistent authenticated
dictionaries based on red-black trees are sketched in some detail,
although the resulting designs are different.  Specifically, in the
former work, although multiple versions of the authenticated
red-black tree are maintained, the collection of versions is itself
not authenticated; the latter work uses a second, non-persistent
authenticated dictionary to authenticate the tree versions.

\section{\label{sec:Timelines}Secure Timelines}

We define a secure timeline within a \emph{service domain}.  A
service domain comprises a system offering a particular
service---the \emph{service} of the domain---and a set of clients
who use that system for that service---the \emph{clients} of the
domain.  Such a service domain could be, for example, the file
server and all clients of a secure file system, or an
enterprise-wide certification authority along with all certificate
subjects within that enterprise.

Within the context of a service domain, a secure timeline is a
tamper-evident, temporally-ordered, append-only sequence of the
states taken by the service of that domain.  In a sense, a secure
timeline defines an authenticated logical clock for the service.
Each time step of the clock is annotated with the state in which the
service is at the time, and an authenticator.  The authenticator is
tamper-evident: given the authenticator of the latest time step of
the timeline, it is intractable for the service or for any other
polynomially-bound party to ``change history'' unobtrusively by
altering the annotations or authenticators of past time steps.

In this work, we consider secure timelines based on one-way (second
pre-image-resistant) hash functions.  Assuming, as is common, that
one-way hash functions exist, we use such functions to define the
``arrow of time.''  In other words, given a presumably one-way hash
function $h$ such as SHA-1~\cite{SHA1}, if $b=h(a)$, then we conclude
that value $a$ was known before value $b$, or $a$ precedes $b$,
since given $b$ the probability of guessing the right $a$ is
negligible.

A simple recursive way to define a secure timeline is as follows: if
at logical time $i$ the clock has authenticator $T_i$, then at the
next logical time step $i+1$, the hash function $h$ is applied to
the previous clock authenticator $T_i$ and to the current state of
the system $S_i$.  Assuming that $f$ is a one-way digest function
from system states to digests, then $T_{i+1}=h(i\|T_i\|f(S_i))$,
where $\|$ denotes concatenation.  Given $T_{i+1}$, it is
intractable to produce appropriate $\alpha$ such that
$T_{i+1}=h(i\|{T_i}'\|\alpha)$, so as to make an arbitrary
authenticator ${T_i}'\neq T_i$ appear as the timeline authenticator
of logical step $i$, from the second pre-image resistance of the
hash function.  Similarly, for a given $T_{i+1}$ only a unique state
digest $d_i=f(S_i)$ is probable, and, from the one-way property of
the state digest function $f$, only a unique system state $S_i$ is
probable.  Therefore, authenticator $T_{i+1}$ is, in a sense, a
one-way digest of all preceding authenticators and system states, as
well as of their total temporal ordering.

Many existing accountable services match the secure timeline
paradigm, since secure timelines are a generalization of secure time
stamping services (TSS)~\cite{Haber1991}.  The service state of a
TSS is an authenticated dictionary of all document digests submitted
to it during a time stamping round.  The Key Archival Service (KAS)
by Maniatis and Baker~\cite{Maniatis2002} is another service with a
timeline, where the service state is now a persistent authenticated
dictionary of all certificates and revocation records issued by a
Certification Authority.  Similarly, any service that maintains
one-way digests of its current state can be retrofitted to have a
secure timeline.  Consider, for example, Kocher's Certificate
Revocation Trees (CRT)~\cite{Kocher1998}.  The state of the service
at the end of each publication interval consists of a hash-tree of
all published revocation records.  The root hash of the CRT is a
one-way digest of the database.  Consequently, a secure timeline for
the revocation service can easily follow from the above
construction.

\begin{figure}
\centerline{\includegraphics{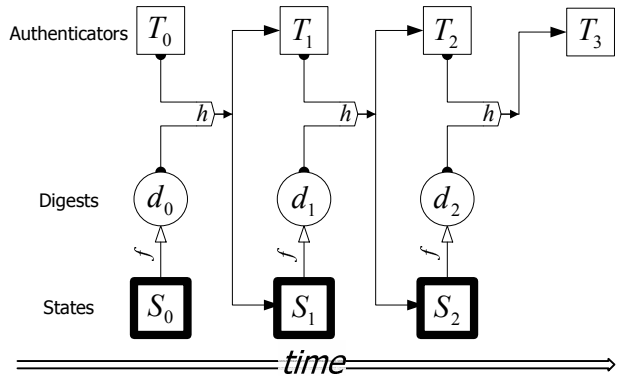}}
\caption{\small The first few steps of a secure timeline.  Time
flows from left to right.  Note that the current authenticator of
the timeline is an input to the next state of the system.  We
explain one way to accomplish this in Section~\ref{sec:RBBTrees}.}
\label{fig:Timeline}
\end{figure}

Figure~\ref{fig:Timeline} illustrates the first few time steps of a
secure timeline.  In the figure, the new timeline authenticator is
also fed into the new state of the system.  Depending on the
definition of the state digest function, a new state of the service
can be shown to be \emph{fresh}, i.e., to have followed the
computation of the authenticator for the previous time step.  In
Time Stamping Services, this places the time stamp of a document
between two rounds of the service.  In the Key Archival Service,
this bounds the time interval during which a change in the
Certification Authority (new certificate, revocation, or refresh)
has occurred.  In a CRT timeline system, this bounds the time when a
revocation database was built.  Some authenticated dictionaries can
be shown to be fresh(e.g., \cite{Buldas1998}), and we explain how we
handle freshness in Section~\ref{sec:RBBTrees}.

Secure timelines can be used to answer two basic kinds of questions:
\emph{existence questions} and \emph{temporal precedence} questions.
Existence questions are of the form ``is $S$ the $i$-th system
state?''  Existence questions can be used to establish that the
service exhibited a certain kind of behavior at a particular phase
in its history.  In the time stamping example, an existence question
could be ``is $d$ the round hash at time $i$?''  A positive answer
allows a client to verify the validity of a time stamp from round
$i$, since time stamps from round $i$ are authenticated with the
root hash of that round.  Temporal precedence questions are of the
form ``did state $S$ occur before state $S'$?''.  In time stamping,
answers to precedence questions can establish precedence between two
time stamped documents.

Answers to both existence and temporal precedence questions are
provable.  Given the last authenticator in the timeline, to prove
the existence of a state in the timeline's past I have to produce a
\emph{one-way path}---a sequence of applications of one-way
functions---from that state to the current timeline authenticator.
Similarly, to prove that state $S$ precedes state $S'$, I have to
show that there exists a one-way path from state $S$ to state $S'$,
which means that the former must precede the latter.  For example,
in Figure~\ref{fig:Timeline}, the path from $S_0$ to $T_1$ to $S_2$
is one-way and establishes that state $S_0$ occurred before $S_2$.
Extending this path to $T_3$ provides an existence proof for state
$S_0$, if the verifier knows that $T_3$ is the latest timeline
authenticator.

Secure timelines are a general mechanism for \emph{temporal
authentication}.  As with any other authentication mechanism,
timeline proofs are useful only if the authenticator against which
they are validated is itself secure and easily accessible to all
verifiers, i.e., the clients within the service domain.  In other
words, clients must be able to receive securely authenticator tuples
of the form $\langle i,T_i\rangle$ from the service at every time
step, or at coarser intervals.  This assumes that clients have a
means to open authenticated channels to the service.  Furthermore,
there must be a unique tuple for every time step $i$.  Either the
service must be trusted by the clients to maintain a unique
timeline, or the timeline must be periodically ``anchored'' on an
unconditionally trusted write-once publication medium, such as a
paper journal or popular newspaper.  The latter technique is used by
some commercial time stamping services~\cite{Surety}, to reduce the
clients' need to trust the service.

For the remainder of this paper, ``time $i$'' means the state of the
service that is current right after timeline element $i$ has been
published, as well as the physical time period until the publication
of the timeline authenticators for time step $i+1$.  For service
$A$, we denote time $i$ as $\langle A,i\rangle$.  Furthermore, we
denote the $i$-th timeline authenticator of service $A$ as $T_i^A$
and the precedence proof from $A$'s time $i$ to $j$ as
$P_{A,i}^{A,j}$, when multiple services are discussed.

\section{\label{sec:Entanglement}Timeline Entanglement}

In the previous section, we described how a secure timeline can be
used by the clients within a service domain to reason about the
temporal ordering of the states of the service in a provable manner.
In so doing, the clients of the service have access to tamper-proof
historic information about the operation of the service in the past.

However, the timeline of service $A$ does not carry much conviction
before a client who belongs to a different, disjoint service domain
$B$, i.e., a client who does not trust service $A$ or the means by
which it is held accountable.  Consider an example from time
stamping where Alice, a client of TSS $A$, wishes to know when Bob,
a client of another TSS $B$, time stamped a particular document
$\mathcal{D}$.  A time stamping proof that links $\mathcal{D}$ to an
authenticator in $B$'s timeline is not convincing to Alice, since
she has no way to compare temporally time steps in $B$'s timeline to
her own timeline, held by $A$.

This is the void that \emph{timeline entanglement} fills.  Timeline
entanglement creates a provable temporal precedence from a time step
in a secure timeline to a time step in another independent timeline.
Its objective is to allow a group of mutually distrustful service
domains to collaborate towards maintaining a common, tamper-proof
history of their collective timelines that can be verified from the
point of view (i.e., within the trust domain) of any one of the
participants.

In timeline entanglement, each participating service domain
maintains its own secure timeline, but also keeps track of the
timelines of other participants, by incorporating authenticators
from those foreign timelines into its own service state, and
therefore its own timeline.  In a sense, all participants
\emph{enforce} the commitment of the timeline authenticators of
their peers.

In Section~\ref{sec:Fundamentals}, we define timeline entanglement
with illustrative examples and outline its properties.  We then
explore in detail three aspects of timeline entanglement:
\emph{Secure Temporal Mappings} in Section~\ref{sec:Mapping}, the
implications of dishonest timeline maintainers in
Section~\ref{sec:Integrity}, and \emph{Historic Survivability} in
Section~\ref{sec:Survivability}.

\subsection{\label{sec:Fundamentals}Fundamentals}

Timeline entanglement is defined within the context of an
\emph{entangled service set}.  This is a dynamically changing set of
service domains.  Although an entangled service set where all
participating domains offer the same kind of service is
conceivable---such as, for example, a set of time stamping
services---we envision many different service types, time stamping
services, certification authorities, historic records services,
etc., participating in the same entangled set.  We assume that all
participating services know the current membership of the entangled
service set, although inconsistencies in this knowledge among
services does not hurt the security of our constructs below.  We
also assume that members of the service set can identify and
authenticate each other, either through the use of a common public
key infrastructure, or through direct out-of-band key exchanges.

Every participating service defines an independent sampling method
to select a relatively small subset of its logical time steps for
entanglement.  For example, a participant can choose to entangle
every $n$-th time step, although other sampling methods are
certainly not precluded.  At every time step picked for
entanglement, the participant sends an authenticated message that
contains its signed logical time and timeline authenticator to all
other participants in the entangled service set.  This message is
called a \emph{timeline thread}.  A timeline thread sent from $A$ at
time $\langle A,i\rangle$ is denoted as $t_i^A$ and has the form
$[A,i,T^A_i,\sigma_A\{A,i,T_i^A\}]$. $\sigma_A\{X\}$ represents
$A$'s signature on message $X$.

\begin{figure}
\centerline{\includegraphics{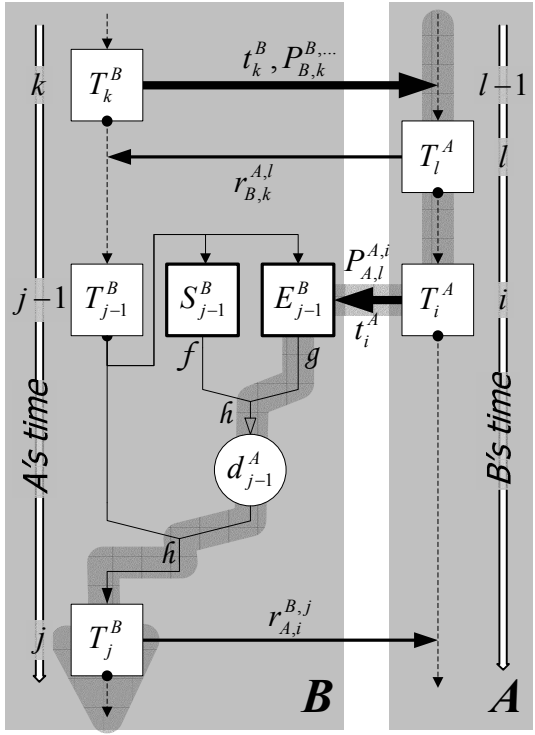}}
\caption{\small Entanglement exchanges between participants $A$ and
$B$.  The workings of $B$ are shown in detail.  We show two
entanglement exchanges, one of time $\langle B, k\rangle$ with time
$\langle A, l\rangle$, and one of time $\langle A, i\rangle$ with
time $\langle B, j\rangle$.  Thick black horizontal arrows show
timeline thread messages.  Thin black horizontal arrows show
entanglement receipt messages.  Vertical black arrows show one-way
operations.  The thick shadowed arrow shows the temporal ordering
effected by thread $t_i^A$ and its receipt $r_{A,i}^{B,j}$.}
\label{fig:Entanglement}
\end{figure}

When participant $B$ receives a correctly signed timeline thread
from participant $A$, it verifies the consistency of that thread
with its local view of collective history and then archives it.
Thread $t_i^A$ is consistent with $B$'s local view of collective
history if it can be proved to be on the same one-way path (hash
chain) as the last timeline authenticator of $A$ that $B$ knows
about (see Figure~\ref{fig:Entanglement}).  Towards this goal, $A$
includes the necessary temporal precedence proof, as described in
Section~\ref{sec:Timelines}, along with the thread that it sends to
$B$.  In the figure, when thread $t_i^A$ reaches $B$, the most
recent timeline authenticator of $A$ that $B$ knows is $T_l^A$.
Along with the thread, $A$ sends the precedence proof
$P_{A,l}^{A,i}$ from its time $\langle A,l\rangle$ to time $\langle
A,i\rangle$.  As a result, $B$ can verify that the new thread
carries a ``legitimate'' timeline authenticator from $A$, one
consistent with history.  If everything checks out, $B$ archives the
new timeline authenticator and associated precedence proof in its
local \emph{thread archive}.

A thread archive serves two purposes: first, it maintains a
participant's local knowledge of the history of the entangled
service set.  Specifically, it archives proof that every participant
it knows about maintains a consistent timeline.  It accomplishes
this by simply storing the threads, which are snapshots in the
sender's timeline, and supporting precedence proofs, which connect
these snapshots in a single one-way chain.  The second purpose of
the thread archive is to maintain temporal precedence proofs between
every foreign thread it receives and local timeline steps.  It
accomplishes this by constructing a one-way digest of its contents,
which is then used along with the system state digest, to derive the
next local timeline authenticator; the modified recursive definition
of timeline authenticators is
$T_{i+1}^A=h(i\|T_i^A\|h(f(S_i^A)\|g(E_i^A)))$, where $g$ is the
one-way digest function for the thread archive, and $E_i^A$ is the
current version of the thread archive at time $\langle A,i\rangle$.
See Section~\ref{sec:RBBTrees} for details on a data structure
capable of fulfilling these requirements.

Participant $B$ responds to the newly reported timeline
authenticator with an \emph{entanglement receipt}.  This receipt
proves that the next timeline authenticator that $B$ produces is
influenced partly by the archiving of the thread it just received.
The receipt must convince $A$ of three things: first, that its
thread was archived; second, that the thread was archived in the
latest---``freshest''---version of $B$'s thread archive; and, third,
that this version of the thread archive is the one whose digest is
used to derive the next timeline authenticator that $B$ produces.
As a result, the entanglement receipt $r_{A,i}^{B,j}$ that $B$
returns to $A$ for the entanglement of thread $t_i^A$ consists of
three components: first, a precedence proof $P_{B,k}^{B,j-1}$ from
the last of $B$'s timeline authenticators that $A$ knows about,
$T_k^B$, to $B$'s timeline authenticator right before archiving
$A$'s new thread $T_{j-1}^B$; second, an existence proof for the
timeline thread $t_i^A$ in the latest version $E_{j-1}^B$ of $B$'s
thread archive, as modified after time $\langle B, j-1\rangle$ (the
equivalent of an undeniable attester in~\cite{Buldas2002}---see also
Section~\ref{sec:RBBTrees}); and, third, a one-way derivation of the
next timeline authenticator of $B$ from the new version of the
thread archive, i.e., the one-way digest $f(S_{j-1}^B)$ of the
current system state.  It is now $A$'s turn to check the validity of
the proofs in the entanglement receipt.  If all goes well, $A$
stores the proof of precedence and reported timeline authenticator
from $B$ in its \emph{receipt archive}.  This concludes the
entanglement process from time $\langle A,i\rangle$ to time $\langle
B,j\rangle$.

The receipt archive is similar to the thread archive; it stores
entanglement receipts that the participant receives in response to
its own timeline threads.  However, it is not an authenticated
archive, since it is not necessary to prove to anyone whether and
when a participant stores the receipts received in response to its
own timeline threads.

After the entanglement of time $\langle A,i\rangle$ with time
$\langle B,j\rangle$, both $A$ and $B$ have in their possession
portable temporal precedence proofs ordering $A$'s past before $B$'s
future. Any one-way process at $A$ whose result is included in the
derivation of $T_i^A$ or earlier timeline authenticators at $A$ can
be shown to have completed before any one-way process at $B$ that
includes in its inputs $T_j^B$ or later timeline authenticators at
$B$.

In this definition of timeline entanglement, a participating service
entangles its timeline at the predetermined sample time steps with
all other services in the entangled service set (we call this
\emph{all-to-all entanglement}).  In this work we limit the
discussion to all-to-all entanglement only, but we describe a more
restricted, and consequently less expensive, entanglement model in
future work (Section~\ref{sec:FutureWork}).

The primary benefit of timeline entanglement is its support for
\emph{secure temporal mapping}.  A client in one service domain can
use temporal information maintained in a remote service domain that
he does not trust, by mapping that information onto his own service
domain.  This mapping results in some loss of temporal
resolution---for example, a time instant maps to a positive-length
time interval.  We describe secure temporal mapping in
Section~\ref{sec:Mapping}.

Timeline entanglement is a sound method of expanding temporal
precedence proofs outside a service domain; it does not prove
incorrect precedences.  However it is not complete, that is, there
are some precedences it cannot prove.  For example, it is possible
for a dishonest service to maintain clandestinely two timelines,
essentially ``hiding'' the commitment of some of its system states
from some members of the entangled service set.  We explore the
implications of such behavior in Section~\ref{sec:Integrity}.

Finally, we consider the survivability characteristics of temporal
proofs beyond the lifetime of the associated timeline, in
Section~\ref{sec:Survivability}.

\subsection{\label{sec:Mapping}Secure Temporal Mapping}

Temporal mapping allows a participating service $A$ to map onto its
own timeline a time step $\langle B,i\rangle$ from the timeline of
another participant $B$.  This mapping is denoted by $\langle
B,i\rangle\mapsto A$.  Since $A$ and $B$ do not trust each other,
the mapping must be secure; this means it should be practically
impossible for $B$ to prove to $A$ that $(\langle B,i\rangle\mapsto
A)=[\langle A,j\rangle,\langle A,k\rangle]$, if $\langle B,i\rangle$
occurred before $\langle A,j\rangle$ or after $\langle A,i\rangle$.

\begin{figure}
\centerline{\includegraphics{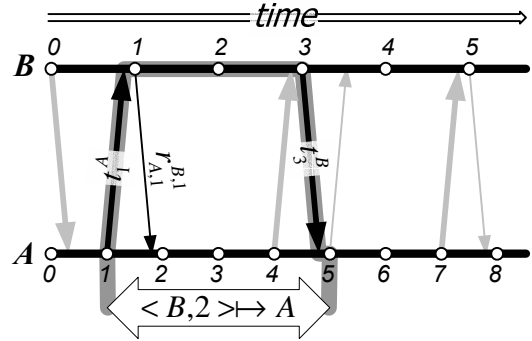}}
\caption{\small Secure mapping of time $\langle B,2\rangle$ onto the
timeline of $A$.  Thick arrows indicate timeline threads.  Thin
arrows indicate entanglement receipts (only the relevant
entanglement receipts are shown). Irrelevant thread and receipt
messages are grayed-out.  The dark broken line illustrates the
progression of values that secure the correctness of the mapping.}
\label{fig:Mapping}
\end{figure}

Figure~\ref{fig:Mapping} illustrates the secure temporal mapping
$\langle B,2\rangle\mapsto A$.  To compute the mapping, $A$ requires
only local information from its thread and receipt archives.  First,
it searches in its receipt archive for the latest entanglement
receipt that $B$ sent back before or at time $\langle B,2\rangle$,
receipt $r_{A,1}^{B,1}$ in the example.  As described in
Section~\ref{sec:Entanglement}, this receipt proves to $A$ that its
time $\langle A,1\rangle$ occurred before $B$'s time $\langle
B,1\rangle$.

Then, $A$ searches in its thread archive for the earliest thread
that $B$ sent it after or at time $\langle B,2\rangle$, which is
thread $t_3^B$ in the example.  This thread proves to $A$ that its
time $\langle A,5\rangle$ occurred after time $\langle B,3\rangle$.
Recall, also, that when $A$ received $t_3^B$ in the first place, it
had also received a temporal precedence proof from $\langle
B,1\rangle$ to $\langle B,3\rangle$, which in the straightforward
hash chain case, also includes the system state digest for $\langle
B,2\rangle$.  Now $A$ has enough information to conclude that
$(\langle B,2\rangle\mapsto A)=[\langle A,1\rangle,\langle
A,5\rangle]$.

Since $A$ has no reason to believe that $B$ maintains its timeline
in regular intervals, there is no more that $A$ can assume about the
temporal placement of state $S_2^B$ within the interval $[\langle
A,1\rangle,\langle A,5\rangle]$.  This results in a \emph{loss of
temporal resolution}; in the figure, this loss is illustrated as the
difference between the length on $B$'s timeline from $\langle
B,2\rangle$ to $\langle B,3\rangle$ (i.e., the ``duration'' of time
step $\langle B,2\rangle$) and the length of the segment on $A$'s
timeline from $\langle A,1\rangle$ to $\langle A,5\rangle$ (the
duration of $\langle B,2\rangle\mapsto A$).  This loss is
proportional to the interval between successive thread messages
between $A$ and $B$.  It can be made shorter, but only at the cost
of increasing the frequency with which $A$ and $B$ send threads to
each other, which translates to more messages and more computation
at $A$ and $B$.  We explore this trade-off in
Section~\ref{sec:Evaluation}.

Secure time mapping allows clients within a service domain to
determine with certainty the temporal ordering between states on
their own service and on remote, untrusted service domains.  Going
back to the time stamping example, assume that Alice has in her
possession a time stamp for document $\mathcal{C}$ in her own
service domain $A$, which links it to local time $\langle
A,0\rangle$, and she has been presented by Bob with a time stamp on
document $\mathcal{D}$ in Bob's service domain $B$, which links
Bob's document to time $\langle B,2\rangle$.  Alice can request from
$A$ the time mapping $\langle B,j\rangle\mapsto A$, shown above to
be $[\langle A,1\rangle,\langle A,5\rangle]$.  With this
information, Alice can be convinced that her document $\mathcal{C}$
was time stamped before Bob's document $\mathcal{D}$ was, regardless
of whether or not Alice trusts Bob or $B$.

In the general case, not all time steps in one timeline map readily
to another timeline.  To reduce the length of temporal precedence
proofs, we use hash skip lists (Section~\ref{sec:Skiplists}) instead
of straightforward hash chains in Timeweave, our prototype.
Temporal precedence proofs on skip lists are shorter because they do
not contain every timeline authenticator from the source to the
destination.  In timelines implemented in this manner, only time
steps included in the skip list proof can be mapped without the
cooperation of the remote service.  For other mappings, the remote
service must supply additional, more detailed precedence proofs,
connecting the time authenticator in question to the time
authenticators that the requester knows about.

\subsection{\label{sec:Integrity}Historic Integrity}

Timeline entanglement is intended as an artificial enlargement of
the class of usable, temporal orderings that clients within a
service domain can determine undeniably.  Without entanglement, a
client can determine the provable ordering of events only on the
local timeline.  With entanglement, one-way paths are created that
anchor time segments from remote, untrusted timelines onto the local
timeline.

However, the one-way properties of the digest and hash functions
used make timelines secure only as long as everybody is referring to
the same, single timeline.  If, instead, a dishonest service $A$
maintains clandestinely two or more timelines or branches of the
same timeline, publishing different timeline authenticators to
different subsets of its users, then that service can, in a sense,
revise history.  Just~\cite{Just1998} identified such an attack
against early time stamping services.  Within a service domain, this
attack can be foiled by enforcing that the service periodically
commits its timeline on a write-once, widely published medium, such
as a local newspaper or paper journal.  In such a way, when there is
doubt, a local client can always request a precedence proof that
links the current timeline authenticator to the most recent widely
published authenticator.

\begin{figure}
\centerline{\includegraphics{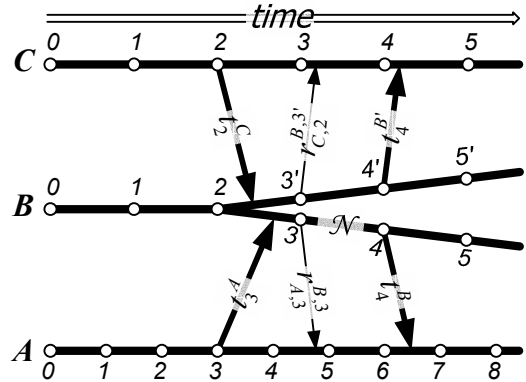}}
\caption{\small An example showing a dishonest service $B$ that
maintains two timelines, entangling one with $A$ and another with
$C$.  Event $\mathcal{N}$ is committed on the bottom branch of $B$'s
timeline, but does not appear on the top branch.}
\label{fig:Integrity}
\end{figure}

Unfortunately, a similar attack can be mounted against the integrity
of collective history, in an entangled service set.  Entanglement,
as described in Section~\ref{sec:Entanglement}, does not verify that
samples from $B$'s timeline that are archived at $A$ and $C$ are
identical.  If $B$ is malicious, it can report authenticators from
one chain to $A$ and from another to $C$, undetected (see
Figure~\ref{fig:Integrity}).  In the general case, this does not
dilute the usability of entanglement among honest service domains.
Instead, it renders unprovable some interactions between honest and
dishonest service domains.  More importantly, attacks by a service
against the integrity of its own timeline can only make external
temporal precedence information involving that timeline
inconclusive; such attacks cannot change the temporal ordering
between time steps on honest and dishonest timelines.  Ultimately,
it is solely the clients of a dishonest service who suffer the
consequences.

Consider, for instance, the scenario of Figure~\ref{fig:Integrity}.
Dishonest service $B$ has branched off its originally unique
timeline into two separate timelines at its time $\langle
B,2\rangle$.  It uses the top branch, with times $3'$, $4'$, etc.,
in its entanglements with service $C$, and its bottom branch, with
times $3$, $4$, etc., in its entanglements with service $A$.  From
$A$'s point of view, event $\mathcal{N}$ is incorporated in $B$'s
state and corresponding timeline at time $\langle B,3\rangle$.  From
$C$'s point of view, however, event $\mathcal{N}$ seems never to
have happened.  Since $\mathcal{N}$ does not appear in the branch of
$B$'s timeline that is visible to $C$, $C$'s clients cannot
conclusively place event $\mathcal{N}$ in time at all.  Therefore,
it is only the client of $B$ who is responsible for event
$\mathcal{N}$ who suffers from this discrepancy.  $C$ does not know
about it at all, and $A$ knows its correct relative temporal
position.

We describe briefly a method for enforcing timeline uniqueness
within an entangled service set in Section~\ref{sec:FutureWork}.

\subsection{\label{sec:Survivability}Historic Survivability}

Historic survivability in the context of an entangled set of
services is the decoupling of the verifiability of existence and
temporal precedence proofs within a timeline from the fate of the
maintainer of that timeline.

Temporal proofs are inherently survivable because of their
dependence on well-known, one-way constructs.  For example, a hash
chain consisting of multiple applications of SHA-1 certainly proves
that the result of the chain temporally followed the input to the
chain.  However, this survivability is moot, if the timeline
authenticators that the proof orders undeniably can no longer be
interpreted, or associated with a real time frame.

Fortunately, secure temporal mapping allows a client within a
service domain to fortify a temporal proof that he cares about
against the passing of the local service.  The client can accomplish
this by participating in more service domains than one; then, he can
proactively map the temporal proofs he cares about from their source
timeline onto all the timelines of the service domains in which he
belongs.  In this manner, even if all but one of the services he is
associated with become unavailable or go out of business, the client
may still associate his proofs with a live timeline in the surviving
service domain.

\begin{figure}
\centerline{\includegraphics{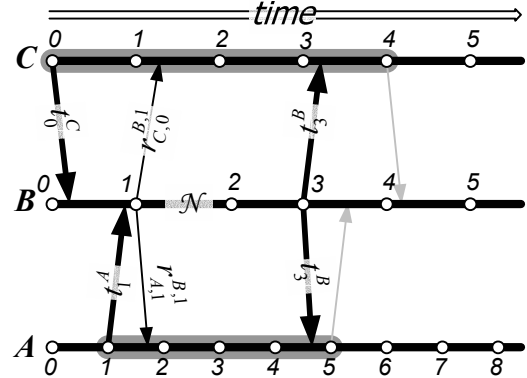}}
\caption{\small An example of mapping event $\mathcal{N}$ onto two
other timelines, to obtain a survivable proof of its temporal
position.  The top shaded line represents $(\mathcal{N}\mapsto A)$
and the bottom shaded line represents $(\mathcal{N}\mapsto C)$.}
\label{fig:MultipleMappings}
\end{figure}

Consider, for example, the scenario illustrated in
Figure~\ref{fig:MultipleMappings}.  David, who belongs to all three
service domains $A$, $B$ and $C$, wishes to fortify event
$\mathcal{N}$ so as to be able to place it in time, even if service
$B$ is no longer available.  He maps the event onto the timelines of
$A$ and $C$---``mapping an event $\mathcal{N}$'' is equivalent to
mapping the timeline time step in whose system state event
$\mathcal{N}$ is included, that is, $\langle B,2\rangle$ in the
example.  Even though the event occurred in $B$'s timeline, David
can still reason about its relative position in time, albeit with
some loss of resolution, in both the service domains of $A$ and $C$,
long after $B$ is gone.  In a sense, David ``hedges his bets'' among
multiple services, hoping that one of them survives.  The use of
temporal mapping in this context is similar in scope to the
techniques used by Ansper et al.~\cite{Ansper2001} for
fault-tolerant time stamping services, although it assumes far less
mutual trust among the different service domains.

\section{\label{sec:Implementation}Implementation}

We have devised two new, to our knowledge, disk-oriented data
structures for the implementation of Timeweave, our timeline
entanglement prototype.  In Section~\ref{sec:Skiplists}, we present
authenticated append-only skip lists. These are an efficient
optimization of traditional hash chains and yield precedence proofs
with size proportional to the square logarithm of the total elements
in the list, as opposed to linear.  In Section~\ref{sec:RBBTrees},
we present RBB-Trees, our disk-based, persistent authenticated
dictionaries based on authenticated search trees.  RBB-Trees scale
to larger sizes than current in-memory persistent authenticated
dictionaries, while making efficient use of the disk.  Finally, in
Section~\ref{sec:Timeweave}, we outline how Timeweave operates.

\subsection{\label{sec:Skiplists}Authenticated Append-only
Skip Lists}

Our basic tool for maintaining an efficient secure timeline is the
authenticated append-only skip list.  The authenticated append-only
skip list is a modification of the simplistic hash chain described
in Section~\ref{sec:Timelines} that yields improved access
characteristics and shorter proofs.

Our skip lists are deterministic, as opposed to the randomization
used in most skip lists proposed in the literature~\cite{Pugh1990}.
Unlike the authenticated skip lists introduced by Goodrich et
al.~\cite{Goodrich2001}, our skip lists are append-only, which
obviates the need for commutative hashing.  Every list element has a
numeric identifier that is a counter from the first element of the
list (the first element is element $1$, the tenth element is element
$10$, and so on).  Every element carries a data value and an
authenticator, similarly to what was suggested in
Section~\ref{sec:Timelines} for single-chain timelines.

The skip list consists of multiple parallel hash chains at different
levels of detail, each containing half as many elements as the
previous one.  The basic chain (at \emph{level 0}) links every
element to the authenticator of the one before it, just like simple
hash chains.  The next chain (at level 1) coexists with the level 0
chain, but only contains elements whose numeric identifiers are
multiples of 2, and every element is linked to the element two
positions before it.  Similarly, only elements with numeric
identifiers that are multiples of $2^i$ are contained in the hash
chain of level $i$.  No chains of level $j$ that is higher than
$\log_2 n$ are maintained, if all elements are $n$.

\begin{figure}
\centerline{\includegraphics{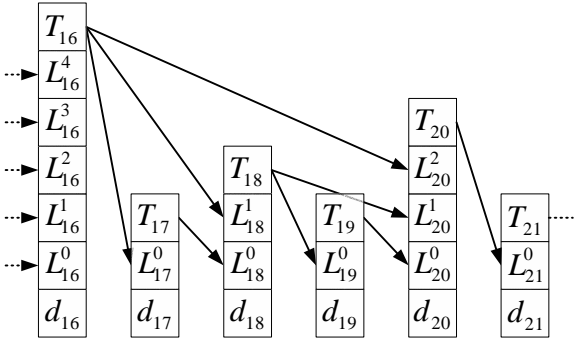}}
\caption{\small Six consecutive skip list elements, element $16$ to
element $21$. Arrows show hash operations from previous
authenticators to links of an element.  The top of each tower is the
resulting authenticator for the element, derived by hashing together
all links underneath it.}
\label{fig:Skiplist}
\end{figure}

The authenticator $T_i$ of element $i$ with data value $d_i$ is
computed from a hash of all the partial authenticators (called
\emph{links}) in each basic hash chain in which it participates---we
share the notation from Section~\ref{sec:Timelines} for data values
and authenticators here.  Element $i$ participates in $l+1$ chains,
where $i=2^lk$, and $2$ does not divide $k$ (in other words, $l$ is
the exponent of $2$ in the factorization of $i$), therefore element
$i$ has the $l+1$ links $L_i^j=h(i,j,d_i,T_{i-2^j})$, $0\leq j\leq
l$, and authenticator $T_i=h(L_i^0\|\ldots\|L_i^l)$.
Figure~\ref{fig:Skiplist} illustrates a portion of such a skip list.
In the implementation, we combine together the element authenticator
with the 0-th level link for odd-numbered elements, since such
elements have a single link, which is sufficient as an authenticator
by itself.

Skip lists allow their efficient traversal from an element $i$ to a
later element $j$ in a logarithmic number of steps: starting from
element $i$, successively higher-level links are utilized until the
``tallest element'' (one with the largest power of 2 in its factors
among all element indices between $i$ and $j$) is reached.
Thereafter, successively lower-level links are traversed until $j$
is reached.  More specifically, an iterative process starts with the
current element $c=i$.  To move closer to the destination element
with index $j$, the highest power $2^l$ of 2 that divides $c$ is
picked, such that $c+2^z\leq j$.  Then element $k=c+2^z$ becomes the
next current element $c$ in the traversal.  The iteration stops when
$c=j$.

The associated temporal precedence proof linking element $i$ before
element $j$ is constructed in a manner similar to the traversal
described above.  At every step, when a jump of length $2^z$ is
taken from the current element $c$ to $k=c+2^z$, the element value
of the new element $d_k$ is appended to the proof, along with all
the associated links of element $k$, except for the link at level
$z$. Link $L_k^z$ is omitted since it can be computed during
verification from the previous authenticator $T_c$ and the new
value $d_k$.

In the example of Figure~\ref{fig:Skiplist}, the path from element
$17$ to element $21$ traverses elements $18$ and $20$.  The
corresponding precedence proof from element $17$ to element $21$ is
$P_{17}^{21}=\{d_{18},L_{18}^1;d_{20},L_{20}^0,L_{20}^2;d_{21}\}$.
With this proof and given the authenticators $T_{17}$ and $T_{21}$
of elements $17$ and $21$ respectively, the verifier can
successively compute
$T_{18}'=h(h(18\|0\|d_{18}\|T_{17})\|L_{18}^{1}))$, then
$T_{20}'=h(L_{20}^0\|h(20\|1\|d_{20}\|T_{18}'\|L_{20}^2))$ and
finally $T_{21}'=h(21\|0\|d_{21}\|T_{20}')$---recall that for all
odd elements $i$, $T_i=L_i^0$.  If the known and the derived values
for the authenticator agree ($T_{21}=T_{21}'$), then the verifier
can be convinced that the authenticator $T_{17}$ preceded the
computation of authenticator $T_{21}$, which is the objective of a
precedence proof, as long as the verifier believes that finding a
second pre-image for the hash function $h$ has negligible
probability.

Thanks to the properties of skip lists, the length of any of these
proofs contains roughly a logarithmic number of authenticators,
links and values in the total number of elements in the timeline.
The worst-case proof for a skip list of $n$ elements traverses
$2\times \log_2(n)$ elements, climbing links of every level between
$0$ and $\log_2(n)$ and back down again, or $\log_2^2(n)$ link
values total.  Assuming that every link and value is a SHA-1 digest
of 160 bits, the worst case proof for a timeline of a billion
elements is no longer than 20 KBytes, and most are much shorter.

Our skip lists are fit for secondary storage.  They are implemented
on memory-mapped files.  Since modifications are expected to be
relatively rare, compared to searches and proof extractions, we
always write changes to the skip list through to the disk
immediately after they are made, to maintain consistency in the face
of machine crashes.  We do not, however, support structural recovery
from disk crashes; we believe that existing file system and
redundant disk array technologies are adequate to prevent and
recover all but the most catastrophic losses of disk bits.

\subsection{\label{sec:RBBTrees}Disk-based Persistent Authenticated
Dictionaries}

This work uses authenticated persistent dictionaries based on trees.
A persistent dictionary maintains multiple versions (or
\emph{snapshots}) of its contents as it is modified.  In addition to
the functionality offered by simple authenticated dictionaries, it
can also provably answer questions of the form ``in snapshot $t$,
was element $d$ in the dictionary?''.  Furthermore, the
authenticated labels for every individual snapshot must also be
maintained.

\begin{figure}
\centerline{\includegraphics{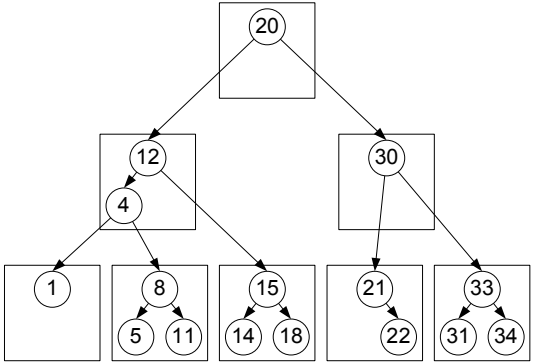}}
\caption{\small An RBB-Tree. Boxes are disk blocks.  In this
example, each non-root disk block contains a minimum of $1$ and a
maximum of $3$ keys.  The authentication labels of the embedded
binary tree nodes are not shown; for any key node, its label
consists of the hash of the label of its left child, its own key,
and the label of its right child, as
in~\cite{Buldas2000,Buldas2002}.  Furthermore, we do not show the
``color'' attribute of the keys in the per-node Red-Black trees,
since they have no bearing in our discussion.}
\label{fig:RBBTree}
\end{figure}

The dictionaries we used in this work can potentially grow very
large, much larger than the sizes of current main memories.
Therefore, we have extended our earlier work on balanced persistent
authenticated search trees~\cite{Maniatis2002} to design on-disk
persistent authenticated dictionaries.  The resulting data
structure, the \emph{RBB-Tree}, is a binary authenticated search
tree~\cite{Buldas2000,Buldas2002} embedded in a persistent
B-Tree~\cite{Bayer1972}\cite[Ch. 18]{CLRv2}.
Figure~\ref{fig:RBBTree} shows a simple RBB-Tree holding 16 numeric
keys.

RBB-Trees, like B-Trees, are designed to organize keys together in
efficient structures that result in few disk accesses per tree
operation.  Every tree node is stored in its own disk block,
contains a minimum of $r-1$ and a maximum of $2r-1$ keys, and has
between $r$ and $2r$ children (the root node is only required to
have between $1$ and $2r-1$ keys).

Unlike traditional B-Trees, RBB-Tree nodes do not store their keys
in a flat array.  Instead, keys within RBB nodes are organized in a
balanced binary tree, specifically a Red-Black
tree~\cite{Bayer1972a}\cite[Ch. 13]{CLRv2}.  We consider RBB-Trees
``virtual'' binary trees, since the in-node binary trees connected
to each other result in a large, piecewise Red-Black tree,
encompassing all keys in the entire dictionary.

It is this ``virtual'' binary tree of keys that is authenticated, in
the sense of the authenticated search trees by Buldas et
al.~\cite{Buldas2000,Buldas2002}.  As such, the security properties
of RBB-Trees are identical to those of authenticated search trees,
including the structure of existence/non-existence proofs.

Since the RBB-Tree is a valid B-Tree, it is efficient in the number
of disk block accesses it requires for the basic tree operations of
insertion, deletion and modification.  Specifically, each of those
operations takes $\mathcal{O}({\log}_r n)$ disk accesses, where $n$
is the total number of keys in the tree. Similarly, since the
internal binary tree in each RBB-Tree node is balanced, the virtual
embedded binary tree is also loosely balanced, and has height
$\mathcal{O}(({\log}_r n)({\log}_2 r))$, that is,
$\mathcal{O}({\log}_2 n)$ but with a higher constant factor than in
a real Red-Black tree.  These two collaborating types of balancing
applied to the virtual binary tree---the first through the blocking
of keys in RBB nodes, and the second through the balancing of the
key nodes inside each RBB node---help keep the length of the
resulting existence/non-existence proofs also bounded to
$\mathcal{O}({\log}_2 n)$ elements.

The internal key structure imposed on RBB-Tree nodes does not
improve the speed of search through the tree over the speed of
search in an equivalent B-Tree, but limits the length of existence
proofs immensely.  The existence proof for a datum inside an
authenticated search tree consists of the search keys of each node
from the sought datum up to the root, along with the labels of the
siblings of each of the ancestors of the sought datum up to the
root.  In a very ``bushy'' tree, as B-Trees are designed to be, this
would mean proofs containing authentication data from a small number
of individual nodes; unfortunately, each individual node's
authentication data consist of roughly $r$ keys and $r$ siblings'
labels.  For example, a straightforwardly implemented authenticated
B-Tree storing a billion SHA-1 digests with $r=100$ yields existence
proofs of length $\lceil{\log}_r 10^9\rceil\times(r\times(160 +
160))$ bits, or roughly 160 KBits.  The equivalent Red-Black tree
yields existence proofs of no more than $2\times\lceil{\log}_2
10^9\rceil\times(160+160)$ bits, or about 18 KBits.  RBB-Trees seek
to trade off the low disk access costs of B-Trees with the short
proof lengths of Red-Black trees.  The equivalent RBB-Tree of one
billion SHA-1 digests yields proofs no longer than
\[\overbrace{\lceil{\log}_r 10^9\rceil}^{\mathrm{B-Tree\ height}}\times
\overbrace{2\times\lceil{\log}_2 r\rceil}^{\mathrm{max\ RB-Tree\
height}}\times\overbrace{(160+160)}^{\mathrm{key}+\mathrm{label}}\] bits
or roughly 22 KBits, with disk access costs identical to those of the
equivalent B-Tree.

\begin{figure}
\centerline{\includegraphics{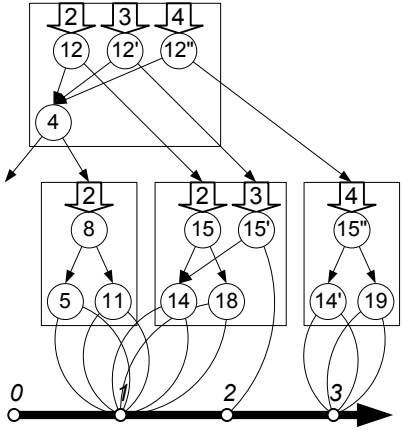}}
\caption{\small A detail from the tree of Figure~\ref{fig:RBBTree}
illustrating dynamic set persistence.  The hollow arrows inside the
RBB nodes denote the different version slots of each node (up to
three in this example).  Version 2 is identical to that of the
original tree shown before.  Version 3 is version 2 after key 18 is
removed.  As a result, a new key node 15' is created and, similarly,
a new key node 12' is created. Version 4 is version 3 after key 19
is inserted.  The RBB node previously holding 14 and 15' had no more
room for key nodes, so a new RBB node was created to hold the new
key nodes 14', 15" and 19.  At the bottom, the snapshot root skip
list is shown.  Nodes that change in a snapshot use the skip list
authenticator of the previous snapshot as their NIL value, to
preserve the ``freshness'' of the snapshot.}
\label{fig:Persistent}
\end{figure}

We have designed dynamic set persistence~\cite[p. 294]{CLRv2} at the
granularity of both the RBB-Node and the embedded key node (see
Figure~\ref{fig:Persistent}).  Each RBB node allows a small number
of persistent versions of itself.  When all the available version
slots or all the available key slots of the RBB-node are exhausted,
a new one is created, containing only the latest key tree snapshot.

The different persistent snapshot roots of the RBB-Tree are held
together in an authenticated linked list---in fact, we use our own
append-only authenticated skip list from
Section~\ref{sec:Skiplists}.

Since each snapshot of the RBB-Tree is a ``virtual'' binary
authenticated search tree, the root label of that tree (i.e., the
label of the root key node of the root RBB node) is a one-way digest
of the snapshot~\cite{Buldas2000,Buldas2002}.  Furthermore, the
authenticated skip list of those snapshot root labels is itself a
one-way digest of the sequence of snapshot roots.  As a result, the
label of the last element of the snapshot root skip list is a
one-way digest of the entire history of operations of the persistent
RBB-Tree.  The snapshot root skip list subsumes the functionality of
the Time Tree in our earlier persistent authenticated Red-Black tree
design~\cite{Maniatis2002}.

We piggy-back the disk location of the snapshot root along with the
snapshot root label as the data value in every skip list element;
however only snapshot root labels participate in skip list
authenticator and link computation, since the particular disk
location of a snapshot root need not be proved to anyone.

In some cases the ``freshness'' of an authenticated dictionary
snapshot has to be provable.  For example, in our description of
secure timelines, we have specified that the system state must
depend on the authenticator of the previous timeline time step.
When the system state is represented by an authenticated dictionary,
an existence proof within that dictionary need not only show that a
sought element is part of the dictionary given the dictionary digest
(root hash), but also that the sought element was added into the
dictionary \emph{after} the authenticator of the previous time step
was known.  As with other authenticated dictionaries, we accomplish
this by making the hash label of NIL pointers equal to the
``freshness'' authenticator, so that all existence proofs of newly
inserted elements---equivalently, non-existence proofs of newly
removed elements---prove that they happened after the given
freshness authenticator was known.  Note that subtrees of the
RBB-Tree that do not change across snapshots retain their old
freshness authenticators.  This is acceptable, since freshness is
only necessary to prove to a client that a requested modification
was just performed (for example, when we produce entanglement
receipts in Section~\ref{sec:Entanglement}), and is required only of
newly removed or inserted dictionary elements.

In standalone RBB-Trees, the freshness authenticator is simply the
last authenticator in the snapshot root list (i.e., the
authenticator that resulted from the insertion of the latest closed
snapshot root into the skip list).  In the RBB-Trees that we use for
thread archives in Timeweave (Section~\ref{sec:Timeweave}), the
freshness authenticator is exactly the authenticator of the previous
timeline time step.

\subsection{\label{sec:Timeweave}Timeweave}

Timeweave is an implementation of the timeline entanglement
mechanisms described in Section~\ref{sec:Entanglement}.  It is built
using our authenticated append-only skip lists
(Section~\ref{sec:Skiplists}) and our on-disk persistent
authenticated search trees (Section~\ref{sec:RBBTrees}).

A Timeweave node maintains four components: first, a service state,
which is application specific, and the one-way digest mechanism
thereof; second, its secure timeline; third, a persistent
authenticated archive of timeline threads received; and, fourth, a
simple archive of entanglement receipts received.

The archive of timeline threads is also an authenticated dictionary,
but persistent in this case.  The data stored in the dictionary are
timeline threads for incoming threads, and timeline thread receipts
for outgoing threads.  The data are ordered by the identity of the
remote service in the thread operation, and then by the foreign
logical time associated with the thread operation.  The dictionary
is implemented as an RBB-Tree and has a well-defined mechanism for
calculating its one-way digest, described in
Section~\ref{sec:RBBTrees}.

Finally, the timeline itself is stored as an append-only
authenticated skip list.  The system digest used to derive the
timeline authenticator at every logical time step is a hash of the
concatenation of the service state digest and the digest of the
thread archive after any incoming and outgoing threads have been
recorded.

The main operational loop of a Timeweave machine is as follows:
\begin{enumerate}
\item Handle client requests and update system state digest $f(S)$.
\item Insert all valid incoming timeline threads into thread archive $E$
and update thread archive digest $g(E)$.
\item Hash together the digests to produce system digest
$d=h(f(S)\|g(E))$.
\item Append $d$ into the timeline skip list, resulting in a new
timeline authenticator $T$, and sign the authenticator.
\item For all incoming timeline threads just archived, construct and
return receipts to thread senders.
\item If it is time to send an outgoing timeline thread, send one to
all peers, and store the receipts in the receipt archive.
\end{enumerate}

The Timeweave machine also allows clients to request local temporal
mappings of remote logical times and temporal precedences between
local times.

\section{\label{sec:Evaluation}Evaluation}

In this section, we evaluate the performance characteristics of
timeline entanglement.  First, in
Section~\ref{sec:DataStructureEvaluation}, we present measurements
from a Java implementation of the Timeweave infrastructure:
authenticated append-only skip lists and RBB-Trees.  Then, in
Section~\ref{sec:TimeweaveEvaluation}, we explore the performance
characteristics of the system as a function of the two basic
Timeweave system parameters, entanglement frequency, and entangled
service set size.

In all measurements, we use a lightly loaded dual Pentium III Xeon
computer at 1 GHz, with 2 GBytes of main memory, running RedHat
Linux 7.2, with the stock 2.4.9-13smp kernel and Sun Microsystems'
JVM 1.3.02.  The three disks used in the experiments are model
MAJ3364MP made by Fujitsu, which offer 10,000 RPMs and 5 ms average
seek time.  We devote 1 GByte of main memory to each experiment, for
in-memory caching and disk block caching alike.  We use a block size
of 16 KBytes, which fixes the order of RBB-Trees to $r=145$.
Finally, for signing we use DSA with SHA-1, with a key size of 1024
bits.

\subsection{\label{sec:DataStructureEvaluation}Data Structure
Performance}

\begin{figure}
\centerline{\includegraphics{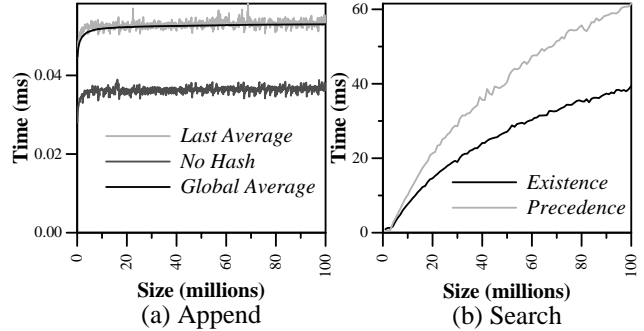}}
\caption{\small Skip list performance.  (a) Append time vs.\ list
size.  ``Global average'' shows average performance over all
operations; ``last average'' shows performance during the last
100,000 operations for a given size; and ``no hash'' shows the same
information as ``last average'' but excludes hashing operations.
(b) Search time vs.\ list size.  ``Existence'' shows searches from a
random element to the end of the list, and ``precedence'' shows
searches between two random elements.}
\label{fig:SkiplistPerformance}
\end{figure}

We measure the raw performance characteristics of our disk-based
authenticated data structures.  Since Timeweave relies heavily on
these two data structures, understanding their performance can help
evaluate the performance limitations of Timeweave.

Figure~\ref{fig:SkiplistPerformance}(a) shows the performance of
skip list appends, for skip list sizes ranging from 100,000 to 100
million elements, in 100,000 element increments.  The figure graphs
average time taken by an append over all operations alongside append
times averaged over the ``last'' few operations (100,000 last
appends for each size plotted).  Late appends cost marginally more
than the average append operation.  The figure also shows the time
taken per append excluding hash operations.  Although more data are
hashed during appends of larger skip lists, the graph shows that the
difference is indistinguishable compared to the setup time of the
Java hashing machinery.

We also measure the performance of skip list searches, in
Figure~\ref{fig:SkiplistPerformance}(b).  We ran 10,000 random
existence search trials per size and 10,000 random precedence search
trials per size.  A random existence search picks a random element
in the list and finds a proof from that element to the end of the
list.  A random precedence search picks two random elements in the
list and finds a proof from the earlier to the later element.
Existence searches are cheaper than precedence searches.  This is
true because random existence proofs for the same list size share,
with high probability, the last few links towards the end of the
list, making better use of cached disk blocks.

\begin{figure}
\centerline{\includegraphics{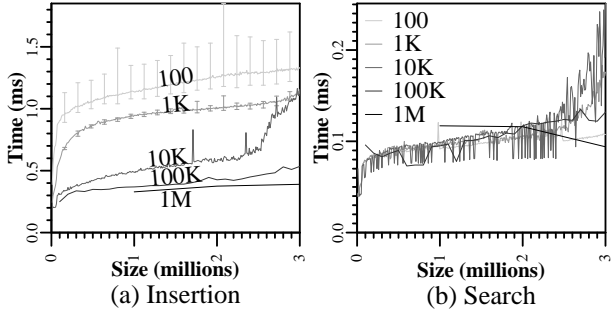}}
\caption{\small RBB-Tree performance for different snapshot sizes.
Curve labels indicate the number of keys per snapshot---from 100
keys to 1 million keys per snapshot.  (a) Insertion time vs.\ tree
size.  (b) Search time vs.\ tree size.}
\label{fig:RBBTreePerformance}
\end{figure}

\begin{table}
\small
\begin{tabular}{r||c|c|c|c|c}
\hline
\hline
Keys per snapshot & 100 & 1K & 10K & 100K & 1M \\
\hline
Tree Size (GB)   & 64 & 45 & 24 & 4 & 0.5 \\
\hline
\hline
\end{tabular}
\caption{\small Tree size on disk as a function of the snapshot size
used to build it.  Sizes shown correspond to trees with three
million keys. }
\label{tab:RBBTreeSize}
\end{table}

We continue by evaluating the performance characteristics of
RBB-Trees.  Figure~\ref{fig:RBBTreePerformance} contains two graphs,
one showing how insertion time grows with tree size
(Figure~\ref{fig:RBBTreePerformance}(a)) and another showing how
search time grows with tree size
(Figure~\ref{fig:RBBTreePerformance}(b)).  Search experiments
consisted of 1,000 random searches for every size increment.  In
both cases, we compare trees with different snapshot sizes, from 100
keys to 1 million keys per snapshot.  Searches are roughly
equivalent across all snapshot sizes, since the ``shape'' of the
tree is independent of how frequently the authentication labels are
computed.  However, different snapshot sizes affect the variance of
search performance.

Smaller snapshot sizes result in more disk blocks stored, because
they create more versions of partially full
blocks. Table~\ref{tab:RBBTreeSize} shows the disk size of a three
million key RBB-Tree with varying snapshot sizes.  This makes
shorter-snapshot trees more expensive during insertion, as can be
seen in Figure~\ref{fig:RBBTreePerformance}(a).

\begin{figure}
\centerline{\includegraphics{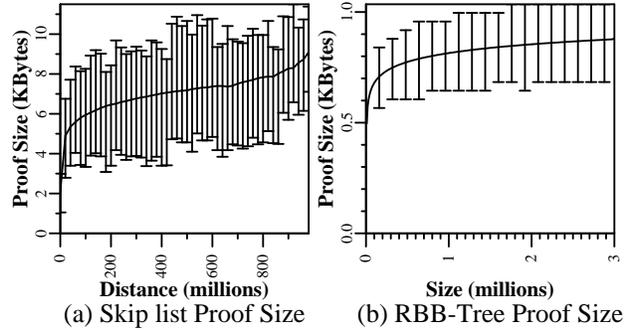}}
\caption{\small Proof sizes (minimum, average, maximum) in skip
lists and RBB-Trees vs.\ structure size.  (a) Proof size vs.\
distance between the proof ends of a skip list proof.  (b) Proof
size vs.\ tree size.}
\label{fig:ProofSizes}
\end{figure}

Finally, we graph proof sizes in skip lists
(Figure~\ref{fig:ProofSizes}(a)) and RBB-Trees
(Figure~\ref{fig:ProofSizes}(b)).  Both graphs show proof sizes in
KBytes, over 10,000 uniform random trials in a skip list of 100
million elements and an RBB-Tree of three million elements,
respectively.  The skip list graph uses the element distance as the
$x$ axis---the distance of elements $i$ and $j$ is $j-i$ elements.
The curve starts out as a regular square logarithmic curve, except
for large distances, close to the size of the entire list.  We
conjecture that the reason for this exception is that for random
trials of distance close to the entire list size, all randomly
chosen proofs are worst-case proofs, including every link of every
level between source and destination, although we must explore this
effect further.  The RBB-Tree graph shows a regular logarithmic
curve.

\subsection{\label{sec:TimeweaveEvaluation}System Performance}

Although microbenchmarks can be helpful in understanding how the
basic blocks of Timeweave perform, they cannot give a complete
picture of how the system performs in action.  For example, very
rarely does a Timeweave machine need to insert thousands of elements
into a skip list back-to-back.  As a result, the disk block caching
available to batched insertions is not available for skip list usage
patterns exhibited by Timeweave.  Similarly, most searches through
timelines only span short distances; for one second-long timeline
time steps with one entanglement process per peer every 10 minutes,
a Timeweave machine barely needs to traverse a difference of
$10\times 60=600$ elements to extract a precedence proof, unlike the
random trials measured in Figure~\ref{fig:SkiplistPerformance}.

In this section we measure the performance of a Timeweave machine in
action.  Assuming that the machine uses timeline steps that last one
second, we measure the amount of time, per second, that the machine
spends on Timeweave maintenance.  Timeweave maintenance consists of
the different steps taken to verify, archive and acknowledge
timeline threads.  By varying the number of thread messages arriving
at the Timeweave machine per time step, we simulate the maintenance
loads resulting from different service set sizes.  For example, if
the entangled service set consists of $1200$ Timeweave machines, and
they all entangle with each other once every $600$ steps (i.e., once
every ten minutes), then every Timeweave machine sends and receives
two threads per time step.  In our experiments, we combine the
receipt message sent by the Timeweave machine with the next thread
message originating from the same machine.  Although the same number
of entanglement processes take place as in the protocol described in
Section~\ref{sec:Entanglement}, some redundant data transmissions
are eliminated.

\begin{figure}
\centerline{\includegraphics{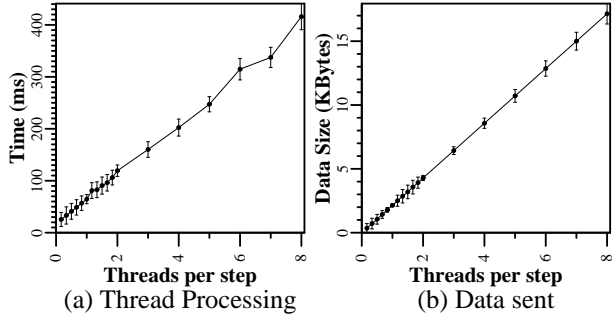}}
\caption{\small Timeweave performance for different thread-per-step
ratios.  The errorbars show one standard deviation around the
average. (a) Insertion time vs.\ tree size.  (b) Search time vs.\
tree size.}
\label{fig:TimeweavePerformance}
\end{figure}

Figure~\ref{fig:TimeweavePerformance}(a) shows the time it takes a
single machine to perform Timeweave maintainance per second-long
time step.  The roughly linear rate at which maintenance processing
grows with the ratio of threads per time step indicates that
all-to-all entanglement can scale to large entangled service sets
only by limiting the entanglement frequency.  However, for
reasonably large service sets, up to $1000$ Timeweave machines for
10-minute entanglement, maintenance ranges between 2 and 8\% of the
processing resources of a PC-grade server.

Figure~\ref{fig:TimeweavePerformance}(b) shows the amount of data
sent per time step from a single Timeweave machine.  Although the
data rate itself is no cause for concern, the number of different
destinations for secure transmissions could also limit how
all-to-all entanglement scales.  Again, for entangled service sets
and entanglement intervals that do not exceed two or three threads
per time step, Timeweave maintenance should not pose a problem to a
low-end server machine with reasonable connectivity.

\section{\label{sec:FutureWork}Conclusion}

In this work we seek to extend the traditional idea of time stamping
into the concept of a secure timeline, a tamper-proof historic
record of the states through which a system passed in its lifetime.
Secure timelines make it possible to reason about the temporal
ordering of system states in a provable manner.  We then proceed to
define timeline entanglement, a technique for creating undeniable
temporal orderings across mutually distrustful service domains.
Finally, we design, describe the implementation of, and evaluate
Timeweave, a prototype implementation of our timeline entanglement
machinery, based on two novel authenticated data structures:
append-only authenticated skip lists and disk-based, persistent
authenticated search trees.  Our measurements indicate that sizes of
a few hundred service domains can be efficiently entangled at a
frequency of once every ten minutes using Timeweave.

Although our constructs preserve the correctness of temporal proofs,
they are not complete, since some events in a dishonest service
domain can be hidden from the timelines with which that domain
entangles (Section~\ref{sec:Integrity}).  We plan to aleviate this
shortcoming by employing a technique reminiscent of the
signed-messages solution to the traditional Byzantine Generals
problem~\cite{Lamport1982}.  Every time service $A$ sends a thread
to peer $B$, it also piggybacks all the signed threads of other
services it has received and archived since the last time it sent a
thread to $B$.  In such a manner, a service will be able to verify
that all members of the entangled service set have received the
same, unique timeline authenticator from every other service that it
has received and archived, thereby verifying global historic
integrity.

We also hope to migrate away from the all-to-all entanglement model,
by employing recently-developped, highly scalable overlay
architectures such as CAN~\cite{Ratnasamy2001} and
Chord~\cite{Stoica2001}.  In this way, a service only entangles its
timeline with its immediate neighbors.  Temporal proofs involving
non-neighboring service domains use \emph{transitive} secure
temporal mapping, over the routing path in the overlay, perhaps
choosing the route of least temporal loss.

Finally, we are working on a large scale distributed historic file
system that enables the automatic maintenance of temporal orderings
among file system operations across the entire system.

\section{\label{sec:Acknowledgments}Acknowledgments}

We thank Dan Boneh for suffering through the early stages of this
work and Hector Garcia-Molina for many helpful comments and pointed
questions.

This work is supported by the Stanford Networking Research Center, by
DARPA (contract N66001-00-C-8015) and by Sonera Corporation.  Petros
Maniatis is supported by a USENIX Scholar Fellowship.

\tiny{
\begin{verbatim}
$Id: Sec2002.tex,v 5.60 2002/02/06 18:44:49 maniatis Exp $
\end{verbatim}
}

\end{document}